# Molecular Mechanics of Chitin-Protein Interface

Zechuan Yu and Denvid Lau[*]

**Department of Civil and Architectural Engineering**
**City University of Hong Kong**

[*]Corresponding Author
denvid.lau@cityu.edu.hk

*Abstract*: Chitin and protein are two main building blocks for many natural biomaterials such as crustacean cuticles, sponge skeletons and squid beaks. These biological materials gain growing interests with respect to biomimetic material design, since they possess great mechanical properties due to their evolutionarily refined structures at different hierarchy levels. The interaction between chitin and protein critically determines the properties of the composite biological materials. Moreover, living organisms usually encounter a complex environment, and ambient conditions like water, pH and ions are critical factors towards the structural integrity of biomaterials such as the chitin-protein composites. It is therefore essential to study the chitin-protein interface under different environmental conditions. Here, an atomistic model consisting of a chitin substrate and a protein filament is constructed, which is regarded as a representative of the chitin-protein interface existing in many chitin-based biomaterials. Based on this model, the mechanical properties of chitin-protein interface under different moisture and pH values are investigated through molecular dynamics simulations. The results reveal that there is a weakening effect of water towards the chitin-protein interface, as well as the influence of acidity, *i.e.* the protonated protein in acidic environment forms a stronger adhesion to chitin than that in the alkaline environment. In addition, the effect from side-chain of protein is studied and it is found that certain kinds of amino acid can form hydrophobic connections to chitin surface, which means that these peptides partly dodge the weakening effect of water. Our observation indicates that terminuses and side-chains in protein are of crucial importance in forming interfacial hydrogen bonds, which connect protein to chitin and contribute to the adhesion. From our full atomistic models, we can observe some molecular mechanisms about how protein interacts with chitin in different conditions, which may spotlight the engineering on biomaterials with similar interfaces.

# Introduction

Among natural polymers, chitin is a most abundant one, especially in animal kingdom. Chitin fibrils possess good mechanical properties, and they are considered as the loading-bearing scaffolds supporting the fundamental structure of composite material, of which another component is usually protein [1]. The remarkable affinity between chitin and protein makes their combination very common in nature. Lots of biological materials are made of chitin and protein, among which are the exoskeleton of sponges, the cuticles of crustaceans and insects, the beaks of squids and the fangs of spiders, to name a few [2-6]. With respect to biomimetic thinking, these natural materials gain research interests for not only their great mechanical properties (*e.g.* high stiffness and stability), but also their versatile capabilities. The versatility of chitin-based materials is outstanding in animal evolutions and it is achieved via structural rearrangement or additive reinforcing. For example, lobster cuticles have different layers with varied structures, where the outer layer is harder and stiffer than the inner layer. Multi-scale simulations of lobster cuticles elucidate that structural effect as well as mineral content determine the cuticle stiffness [3]. Following this idea, spiders reinforce their fangs via enriching the mineral content. Researchers find that in different regions of fang, different kinds of metal ions may result in different mechanical performance [6]. In contrast, squids tune the mechanical properties of their beaks via controlling other variables. Experimental studies of squid beaks indicate that the gradient in water content and cross-link degree cause the gradient in the material stiffness [7]. According to those studies, it is noted that the mechanical properties of chitin-protein composite are subject to many factors from its structure, composition and ambient environment. With chitin and protein combining together, the interaction between them should have a profound influence on the material properties. Moreover, because the chitin-protein interface sensitive the ambient environment, the interface can be considered as the entry where the environmental factors start to take effect. Environmental factors such as solvent, pH and ions, which often correlate with biological systems, also affect the chitin-protein interface. Through performing molecular dynamics simulations, previous study has revealed that water molecules have a weakening effect on the chitin-protein integrity [8]. This phenomenon is reproduced, from which we have also observed diminished hydrogen bonds between chitin and protein in water. In addition, here we propose a "hydrophobic" conformation of protein with an histidine amino acid placed at the N-terminus, may form a connection to chitin that eliminate the weakening effect from water partially. Also, the effect from pH is reported. The solution pH is able to change the protonation state of titratable groups of protein, and thus modulates the interaction between protein and other components [9]. Different behaviors of chitin and protein are observed in acid and alkaline environments.

# Methods

## Modeling

A chitin substrate and a protein filament are constructed. The conformation of α-chitin has been resolved in the *ab inito* based study [10]. The unit cell of α-chitin is replicated ten times in *x* axis, one in *y* axis and five in *z* axis. Protein is in the form of β-strand consisting of five glycines. The small protein filament is placed over the chitin substrate. Two termini of the protein, i.e., the —$NH_2$ and —COOH, are titratable groups, and they are modified to three different states, which are non-ionic, acid and alkaline. The N-terminus and C-terminus of the protein are different. In acid environment, the N-terminus adopts the —$NH_3^+$ form and the C-terminus is in the form of —COOH, while in the alkaline environment, they are —$NH_2$ and —$COO^-$ respectively. In total, four models are built. Three of them represent non-ionized, acid and alkaline environments and an additional one contains an histidine N-terminus.

**Simulation**
LAMMPS is used to perform the simulations [11]. CHARMM36 force fields for proteins and carbohydrates are adopted to govern the movement of all the models [12,13]. VMD is used for visualization [14]. The equilibration process for every model is stated as follows. The model is first minimized using the conjugate gradient algorithm, which adjusts the atom coordinates in order to reach a minimum-energy status. It is then heated up from 50 K to 300 K in 20 ps. After equilibrated at 300 K for 50 ps in an NVT ensemble, the model is equilibrated at 300 K and 1 atm for 500 ps in an NPT ensemble, following which the model is equilibrated in NVT ensemble for another 1 ns. After achieving equilibration in dry case, the system is then immersed into water, equilibrated in NVT ensemble for 1 ns and saved for later simulations. Steered molecular dynamics simulations are performed to measure the adhesion strength between chitin substrate and protein filament. The time step is 1 fs during the heating process, and 2 fs in the subsequent equilibrations. Figure 1 outlines the overall simulation procedures.

## Results and Discussions

**The Comparison Between Acid and Alkaline Environments**
This section has been included in a publication [Yu, Z., Xu, Z., & Lau, D. (2014). Effect of acidity on chitin–protein interface: a molecular dynamics study. BioNanoScience, 4(3), 207-215] and it is quoted here for presentations in the conference. Free energy change ($\Delta F$) is measured. In acid environment, $\Delta F$ is 5.3 kcal/mol, while the $\Delta F$ in alkaline environment is 3.0 kcal/mol. Result shows a stronger adhesion between chitin and acid-state protein, which might be ascribed to the additional hydrogen atom in C-terminus and the nature of hydrogen bond. A hydrogen bond is composed of a donor and an acceptor [16]. In this study, nitrogen (N) and oxygen (O) atoms are regarded as hydrogen bond acceptors. With an attached hydrogen atom (structures like N-H and O-H), they become hydrogen bond donors. Because of the absence of hydrogen atom in —$COO^-$, it can only act as the hydrogen bond acceptor, while the –COOH can be both donor and acceptor, providing more opportunities to form hydrogen bonds. This additional hydrogen atom therefore strengthens the adhesion between protein filament and chitin surface. However, it is noted that the N-terminus is rooted into the chitin surface in all the concerning cases. This is probably because the two donors in —$NH_2$ already permit sufficient hydrogen bonds, so the improvement from the additional hydrogen atom in —$NH_3^+$ is less significant.

**The Weakening Effect from Water**
The non-ionized protein is attached to chitin substrate in both dry and wet circumstances. These equilibrated dry and wet models are tested via steered molecular dynamics approach as shown in Figure 1d. Free energy is calculated during the peeling process using second-order expansion of Jarzynski's equality [15]. The change of free energy (referred to as $\Delta F$) from attachment state to the detachment state indicates the adhesion strength between chitin and protein. The $\Delta F$ is 7.1 kcal/mol in wet case, which is around one third of the $\Delta F$ in wet case, implying that water molecules cast a weakening effect towards chitin-protein interface. This result is in line with previous study [8].

**The "Hydrophobic" Conformation**
As aforementioned, protein models with two sequences (*i.e*. GGGGG and HGGGG) are built and sent for equilibration with chitin substrate. When attached, protein and chitin form hydrogen bonds, which indicate adhesion strength. We computed the average number of hydrogen bonds formed between protein filament and chitin substrate in dry and wet cases. The criteria to detect hydrogen bonds are 4 Å for the donor-acceptor distance and 35 degree for donor-hydrogen-acceptor angle [8]. Figure 2 shows the conformation of proteins in dry and wet cases, it is observed that water reduces the number of hydrogen bonds. Comparing these two protein conformations, results are that protein with HGGGG sequence is capable of forming more hydrogen bonds with chitin in both dry and wet cases.

## Conclusion

Chitin-protein interface plays a profound role in determining the overall properties of the composite material and may act as the entry where the environmental factors cast effect. Following this idea, we investigate the effects of water and acidity on the interface. Conclusions are that terminuses of protein are important in determining the molecular mechanisms of chitin-protein interface. Through either changing the conformations of terminal part of protein or altering the protonation state of terminuses, water and acidity affect the interfacial strength obviously.

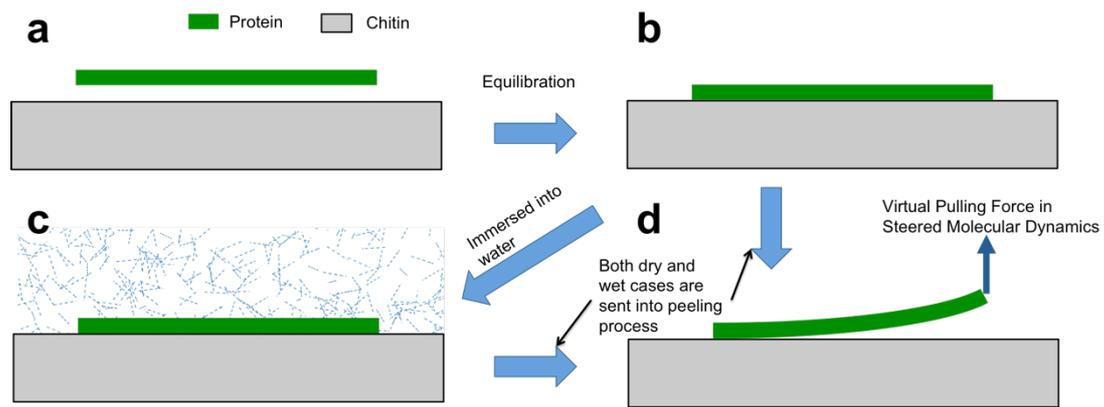

**Figure 1** Flow chart of the simulation process.

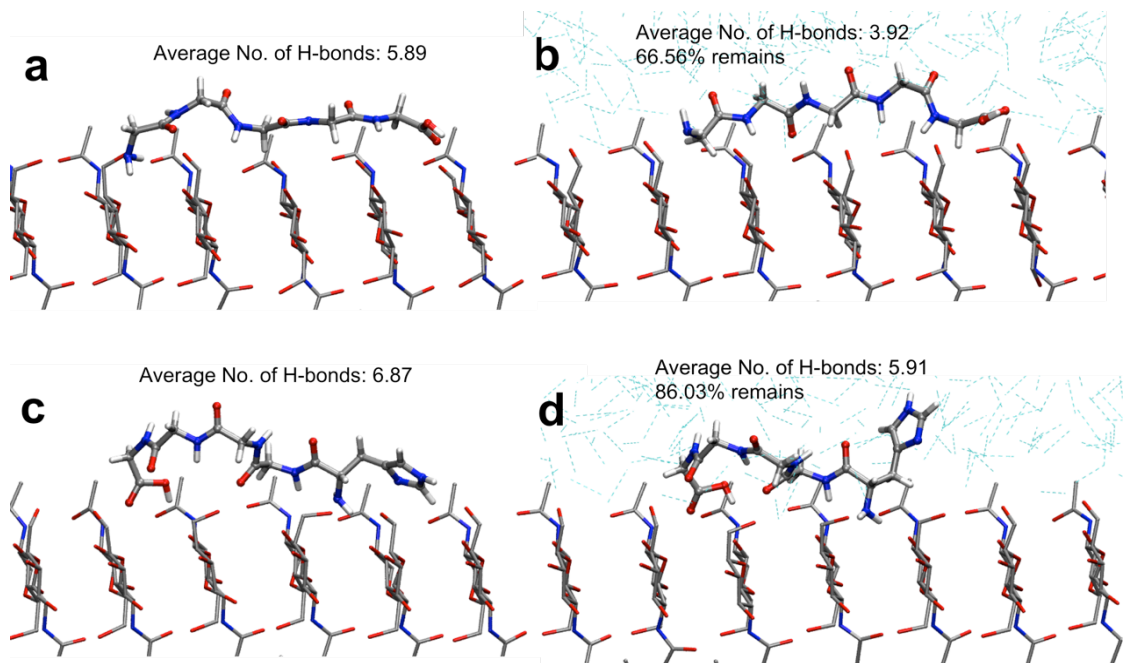

**Figure 2 a b** The protein in sequence GGGGG is attached to chitin in dry and wet conditions.  **c d** The protein in sequence HGGGG is attached to chitin substrate in dry and wet conditions.